\documentclass[a4paper]{IEEEtran}

\usepackage{algorithm}
\usepackage{algpseudocode}
\usepackage{graphicx}
\usepackage{epstopdf}
\usepackage{scalefnt}
\usepackage[table]{xcolor}
\usepackage[cmex10]{amsmath}
\usepackage{etoolbox}
\usepackage[framemethod=TikZ]{mdframed}
\usepackage[noadjust]{cite}
\usepackage{filecontents}

\newtheorem{rmk}{Remark}

\IEEEoverridecommandlockouts
\begin{document}
\title{Frequency of Arrival-based Interference Localization \\ Using a Single Satellite}
\author{
	\IEEEauthorblockN{ Ashkan Kalantari\IEEEauthorrefmark{1}, Sina Maleki\IEEEauthorrefmark{1}, Symeon Chatzinotas\IEEEauthorrefmark{1}, and Bj\"{o}rn Ottersten\IEEEauthorrefmark{1} }
	
	\IEEEauthorblockA{\IEEEauthorrefmark{1}SnT, University of Luxembourg\\ 
	Emails: \{ashkan.kalantari, sina.maleki, symeon.chatzinotas,  bjorn.ottersten\}@uni.lu}
}
\maketitle
\begin{abstract}
Intentional and unintentional interferences are an increasing threat for the satellite communications industry. In this paper, we aim to localize an interference with unknown location using frequency of arrival (FoA) technique by only relying on the measurements obtained through a single satellite. In each time instance, the satellite samples the interference and forwards it to the gateway to estimate its frequency. Since the satellite moves, each estimated frequency includes a Doppler shift, which is related to the location of the unknown interferer. The satellite's position, velocity, oscillator frequency, and the interference frequency are used at the gateway to build a location-related equation between the estimated frequency and the location of the unknown interference. Simultaneously with the interference signal, the satellite samples a reference signal to calibrate the estimated frequency and compensate for the mismatches between the available and real values of the satellite's position, velocity, and oscillator frequency. Multiple location-related equations obtained based on the FoA measurements, (at least two), along with the equation of the earth surface are used to localize the unknown interference. Simulations show that increasing the number of these equations, and the satellite velocity can improve the localization accuracy by $80\%$ and $95\%$, respectively.   
\end{abstract}
\begin{IEEEkeywords}
Doppler shift, frequency of arrival, interference, localization, reference signal.
\end{IEEEkeywords}
\section{Introduction}
In the satellite industry, interference is an important and increasing concern, which can cause considerable revenue loss due to service interruption. Broadly speaking, interference can be divided into two major categories: 1) unintentional, e.g., created by human mistake or equipment mismatch, and 2) intentional, e.g., jammers. In order to effectively handle interference, first, the interferer needs to be localized.

There has been interest toward interference localization in the satellite communication society. The authors of~\cite{Smith:1991} perform time difference of arrival (TDOA) along with phase measurements to localize an unknown interferer using two geostationary (GEO) satellites. In~\cite{Ho:1993}, three out of four satellites exposed to interference are used to derive TDOA measurements for localizing an unknown interferer. The performance of the localization in Eutelsat satellites is presented in~\cite{Haworth:1997} where TDOA and frequency difference of arrival (FDOA) measurements are used to localize an unknown interferer. The altitude constraint in~\cite{Ho:1997,Pattison:2000} is considered to improve the localization accuracy by employing TDOA and/or FDOA technique(s). Two antennas on a spinning satellite in~\cite{Teng:2010} are used to localize an unknown interferer. FDOA measurements done by more than two satellites in~\cite{Jinzhou:2012} are used to localize an interferer. It is shown that, in contrast to TDOA, FDOA accuracy is not affected by the bandwidth of the interference signal.

In addition, there are patents concerning the localization of unknown interferer using satellite. In~\cite{Effland:1991}, one mainly affected GEO satellite and an adjacent GEO satellite capture multiple samples of interference and transmit them to a two-antenna gateway where multiple TDOA and FDOA calculations are done to localize the unknown interferer. In addition, a known reference signal is used to compensate for equipment mismatches. Patents~\cite{Haworth:2000,Rideout:2004} consider a system model similar to~\cite{Effland:1991}, while~\cite{Haworth:2000} works with more inclined satellites, and~\cite{Rideout:2004} can localize an interferer with varying frequency. Patents~\cite{Ho:2010:1,Ho:2010:2} use three satellites for sampling the interference and perform a weighted combination of two TDOA and two FDOA measurements for localization. FDOA technique is employed in~\cite{Ho2013} to localize the unknown interferer using only one affected GEO satellite.

In this work, we consider a satellite that receives uplink signal from a gateway on the earth within the Ka band. At the same time, the satellite receives narrow band uplink interference from an unknown source that is transmitting within the same frequency band as the gateway. Here, we use the term unknown to refer to the location of the interference, which is unknown. Our primary goal is to localize the unknown interferer by applying the FoA technique to the received samples of the interference signal at the gateway that are only sent by the affected satellite, or by a single satellite that is responsible for interference localization. To this end, the satellite samples the interference signal at each time instance and transmits it to the gateway. At the gateway, the frequency of each received sample is estimated. Since the GEO satellite drifts~\cite{Ha:1987} and LEO, MEO, and retrograde GEO\footnote{A retrograde GEO satellite is placed in the GEO orbit and goes against the direction of the Earth rotation.} satellites naturally move, each estimated frequency at the gateway includes a specific amount of Doppler shift which is related to the position of the unknown interferer. Based on this fact, the gateway can build a location-related equation between each estimated frequency and the location of the unknown interferer. Assuming that the frequency of the unknown interferer is known, to build a location-related equation, the gateway requires the frequency of the satellite's down conversion oscillator as well as satellite's positions and velocities when sampling and forwarding the interference signal. However, these values are erroneous. To compensate for errors, a reference signal from a known location on the earth is transmitted to the satellite, and then forwarded to the gateway. The position of the interferer can be derived using at least two location-related equations plus equation of the earth surface. Compared to~\cite{Ho2013}, we use FoA instead of FDOA to localize the unknown interferer, which based on the results in Section~\ref{sec:sim}, improves the localization accuracy. In contrast to~\cite{Ho2013}, we perform more accurate analysis by considering all the terms which cause frequency shift. We use Taylor series approximation along with Newton method to solve the system of location-related equations which are nonlinear. In addition, we propose using on-board interference localization to improve the localization accuracy.

The remainder of the paper is organized as follows. In Section~\ref{sec:sys:mod}, the network configuration is introduced and the frequency shift in interference and reference signals are modeled. In Section~\ref{sec:alg}, system of nonlinear equations is formulated and solved. The simulation results are presented in Section~\ref{sec:sim}, and the conclusions are drawn in Section~\ref{sec:con}.

\emph{Notation}: Upper-case and lower-case bold-faced letters are used to denote matrices and column vectors, respectively. Superscript $(\cdot)^T$ represents transpose and $\|\cdot\|$ is the Frobenius norm.
\section{Signal and System Model}\label{sec:sys:mod}
\begin{figure}[t!]
	\centering
	\includegraphics[width=10cm]{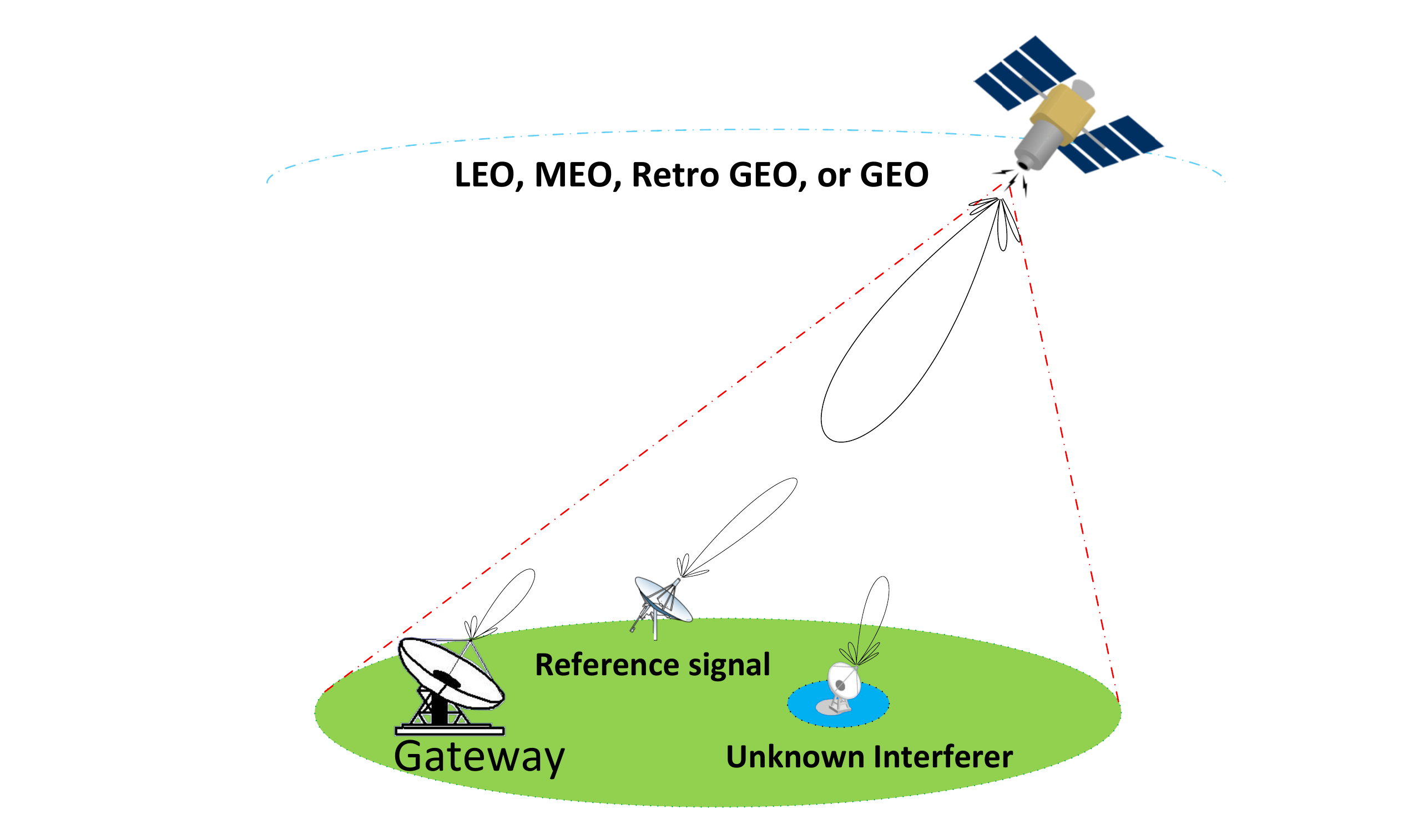}
	\caption{An affected or a localization-dedicated satellite receiving interference and reference signals in uplink.}
	\label{fig:sys:mod}
\end{figure}
We consider a satellite which receives uplink signal from a gateway within the Ka band. Concurrently, the satellite receives narrow band uplink interference from an unknown transmitter within the same frequency band as the uplink signal from the gateway. A reference signal is transmitted to the satellite to compensate for the errors. The whole scenario is summarized in Fig.~\ref{fig:sys:mod}. The central frequency of the interference signal is shown by $f_u$ and since it is interfering with the main uplink signal, we assume that $f_u$ is known. Although $f_u$ may be changed intentionally and/or due to instability of the electronics, for the sake of simplicity, $f_u$ is considered to be fixed through the time. Also, we assume that the derived signal is turned off during sampling the interference signal. All the vectors in this section and Section~\ref{sec:alg} are in Cartesian coordinates. The subscripts \textit{u}, \textit{r}, \textit{s}, \textit{gw}, \textit{ul}, and \textit{dl} are used in the equations instead of the terms: unknown interferer, reference transmitter, satellite, gateway, uplink, and downlink, respectively. 

The frequency of the $n$-th sampled interference by the satellite is
\begin{align}
f_{{n_{u,s}}} = {f_u}\left( {1 + \frac{{{\bf{v}}_{{n_{ul}}}^T{{\bf{k}}_{{n_{u,s}}}}}}{{{c_n}}}} \right),
\label{eqn:s:f}
\end{align}
where ${f_{{n_{u,s}}}}$ is the frequency of the $n$-th sampled interference at the satellite, ${{\bf{v}}_{{n_{ul}}}}$ is the velocity of the satellite when sampling, $c_n$ is the propagation speed of the signal in the space, and ${{{\bf{k}}_{{n_{u,s}}}}}$ is the normalized unit vector pointing from the satellite toward the unknown interferer defined as 
\begin{align}
{{\bf{k}}_{{n_{u,s}}}} = \frac{{{\bf{u}} - {{\bf{s}}_{{n_{ul}}}}}}{{\left\| {{\bf{u}} - {{\bf{s}}_{{n_{ul}}}}} \right\|}},
\label{eqn:k:s:f}
\end{align}
where ${{{\bf{s}}_{{n_{ul}}}}}$ is the position of the satellite during uplink and ${\bf{u}} = \left[ {{u_1},{u_2},{u_3}} \right]$ is the location of the unknown interferer. Afterwards, the satellite down converts ${f_{{n_{u,s}}}}$ into
\begin{align}
{ f_{{n_{u,s}}}} - {f_{{T}}}{\rm{ }} = {f_u}\left( {1 + \frac{{{\bf{v}}_{{n_{ul}}}^T{{\bf{k}}_{{n_{u,s}}}}}}{{{c_n}}}} \right) - {f_{{T}}},
\label{eqn:f:con}
\end{align}
where $f_{T}$ is the amount of the frequency down conversion for the $n$-th sample. Subsequently, the satellite forwards the down converted signal to the gateway. Using~\eqref{eqn:f:con}, the frequency of the received signal at the gateway is
\begin{align}
{f_{{n_{u,g}}}} =& \left( {{f_u} + {f_u}\frac{{{\bf{v}}_{{n_{ul}}}^T{{\bf{k}}_{{n_{u,s}}}}}}{{{c_n}}} - {f_{{T}}}} \right)\left( {1 + \frac{{{\bf{v}}_{{n_{dl}}}^T{{\bf{k}}_{{n_{s,g}}}}}}{{{c_n}}}} \right)
\nonumber\\
=& {f_{{n_{dl}}}} + {f_u}\frac{{{\bf{v}}_{{n_{ul}}}^T{{\bf{k}}_{{n_{u,s}}}}}}{{{c_n}}} + {f_{{n_{dl}}}}\frac{{{\bf{v}}_{{n_{dl}}}^T{{\bf{k}}_{{n_{s,g}}}}}}{{{c_n}}} 
\nonumber\\
&+ {f_u}\frac{{{\bf{v}}_{{n_{ul}}}^T{{\bf{k}}_{{n_{u,s}}}}}}{{{c_n}}}\frac{{{\bf{v}}_{{n_{dl}}}^T{{\bf{k}}_{{n_{s,g}}}}}}{{{c_n}}},
\label{eqn:g:f}
\end{align}
where ${f_{{n_{dl}}}} = {f_u} - {f_{{T}}}$ and ${{\bf{k}}_{{n_{s,g}}}} = \frac{{{{\bf{s}}_{gw}} - {{\bf{s}}_{{n_{dl}}}}}}{{\left\| {{{\bf{s}}_{gw}} - {{\bf{s}}_{{n_{dl}}}}} \right\|}}$ with ${{{\bf{s}}_{gw}}}$ being the position of the gateway and ${{{\bf{s}}_{{n_{dl}}}}}$ being the position of the satellite when forwarding the $n$-th sampled interference to the gateway. The last term in~\eqref{eqn:g:f} is very small compared to the other terms when it comes to GEO satellites with a very slow drift and has been neglected in~\cite{Ho2013}. However, we keep it since its effect increases as the velocity of the satellite goes higher, specially for low earth orbit (LEO), medium earth orbit (MEO) or retro GEO satellites. 

The gateway estimates the frequency of the $n$-th sampled interference after receiving it from the satellite. Due to the movement of the satellite, each estimated frequency includes a specific amount of Doppler shift which relates to the position of the unknown interferer. Hence, a location-related equation can be made between each estimated frequency and the location of the unknown interferer. To this end, the gateway requires satellite's positions and velocities during uplink and downlink of the $n$-th sample, the frequency of the satellite's down conversion oscillator, and the frequency of the interference signal while it is being emitted. However, the values related to the oscillator frequency, positions, and velocities are different from their real values due to equipment impairments. To compensate for these errors, the gateway needs to calibrate the estimated frequency of the $n$-th sample. For this purpose, a reference signal from a known location on the earth can be transmitted to the satellite and then forwarded to the gateway in one of the following approaches:
\begin{enumerate}
\item The reference signal is uplinked in the same frequency as the interference signal after a delay. Due to the delay, the reference and interference signals experience different mismatches.
\item The reference signal is uplinked in a different frequency from the interference signal and the satellite samples the interference and reference signals simultaneously.  
\end{enumerate}
Here, the second approach is followed to transmit the reference signal. By following a similar procedure as in~\eqref{eqn:s:f} to~\eqref{eqn:g:f}, the frequency of the $n$-th sample of the reference signal at the gateway is obtained by
\begin{align}
{f_{{n_{r,g}}}} =& \left( {{f_{{r}}} + {f_{{r}}}\frac{{{\bf{v}}_{{n_{ul}}}^T{{\bf{k}}_{{n_{r,s}}}}}}{{{c_n}}} - {f_{{T}}}} \right)\left( {1 + \frac{{{\bf{v}}_{{n_{dl}}}^T{{\bf{k}}_{{n_{s,g}}}}}}{{{c_n}}}} \right)
\nonumber\\
=& {f^{'}_{{n_{dl}}}} + {f_{{r}}}\frac{{{\bf{v}}_{{n_{ul}}}^T{{\bf{k}}_{{n_{r,s}}}}}}{{{c_n}}} + {f^{'}_{{n_{dl}}}}\frac{{{\bf{v}}_{{n_{dl}}}^T{{\bf{k}}_{{n_{s,g}}}}}}{{{c_n}}} 
\nonumber\\
&+ {f_{{r}}}\frac{{{\bf{v}}_{{n_{ul}}}^T{{\bf{k}}_{{n_{r,s}}}}}}{{{c_n}}}\frac{{{\bf{v}}_{{n_{dl}}}^T{{\bf{k}}_{{n_{s,g}}}}}}{{{c_n}}},
\label{eqn:f:r}
\end{align}
where ${f_{{n_{r,g }}}}$ is the estimated frequency of the reference signal at the gateway and ${{{\bf{k}}_{{n_{r,s}}}}}$ is the normalized unit vector pointing from the satellite toward the reference transmitter defined as ${{\bf{k}}_{{n_{r,s}}}} = \frac{{{\bf{r}} - {{\bf{s}}_{{n_{ul}}}}}}{{\left\| {{\bf{r}} - {{\bf{s}}_{{n_{ul}}}}} \right\|}}$ with $\bf{r}$ being the location of the reference transmitter. Next, the gateway calculates the expected frequency of the reference signal using the available erroneous data as
\begin{align}
{f_{{n_{r,g,exp}}}} =& {f_{{r}}} - {f_{{n_{{T_e}}}}} + {f_{{r}}}\frac{{{\bf{v}}_{{n_{u{l_e}}}}^T{{\bf{k}}_{{n_{{{\left( {r,s} \right)}_e}}}}}}}{{{c_n}}} 
+ {f_{{n_{d{l_e}}}}}\frac{{{\bf{v}}_{{n_{d{l_e}}}}^T{{\bf{k}}_{{n_{{{\left( {s,g} \right)}_e}}}}}}}{{{c_n}}} 
\nonumber\\
&+ {f_{{r}}}\frac{{{\bf{v}}_{{n_{u{l_e}}}}^T{{\bf{k}}_{{n_{{{\left( {r,s} \right)}_e}}}}}}}{{{c_n}}}\frac{{{\bf{v}}_{{n_{d{l_e}}}}^T{{\bf{k}}_{{n_{{{\left( {s,g} \right)}_e}}}}}}}{{{c_n}}},
\label{eqn:ref:exp}
\end{align}
where ${f_{{n_{r,g,exp}}}}$ is the expected frequency of the $n$-th sampled reference signal at the gateway, ${{\bf{k}}_{{n_{{{(r,s)}_e}}}}} = \frac{{{\bf{r}} - {{\bf{s}}_{{n_{u{l_e}}}}}}}{{\left\| {{\bf{r}} - {{\bf{s}}_{{n_{u{l_e}}}}}} \right\|}}$, and ${{\bf{k}}_{{n_{{{(s,g)}_e}}}}} = \frac{{{{\bf{s}}_{gw}} - {{\bf{s}}_{{n_{d{l_e}}}}}}}{{\left\| {{{\bf{s}}_{gw}} - {{\bf{s}}_{{n_{d{l_e}}}}}} \right\|}}$. The frequency mismatch for the $n$-th sample is derived using~\eqref{eqn:f:r} and~\eqref{eqn:ref:exp} as
\begin{align}
{\delta _n} = \frac{{{f_u}}}{{{f_{{r}}}}}\left[ {{f_{{n_{r,g}}}} - {f_{{n_{r,g,exp}}}}} \right],
\label{eqn:cal:fac}
\end{align}
where $\delta_n$ is the amount of the frequency mismatch, the factor $\frac{{{f_u}}}{{{f_{{r}}}}}$ is used to convert the frequency of the reference signal into the frequency of the unknown emitter since the reference signal has a different frequency and undergoes a different amount of mismatches. Using~\eqref{eqn:cal:fac}, the calibrated frequency of the $n$-th received interference at the gateway is obtained by ${\widetilde f_{{n_{u,g}}}} = {f_{{n_{u,g}}}} - {\delta _n}$ 
where ${\widetilde f_{{n_{u,g}}}}$ is the calibrated frequency. The difference in the location of the unknown and the reference transmitters leads into different values for ${{\bf{k}}}_{u,s}$ and ${{\bf{k}}}_{r,s}$. Hence, the satellite velocity in the uplink will have different values and errors in the directions of ${{\bf{k}}}_{u,s}$ and ${{\bf{k}}}_{r,s}$, which means that the reference signal does not go through the same amount of mismatches as the unknown interference signal. To improve this, we can perform iterative localization and choose a closer reference transmitter to the unknown interferer after each localization step. After calibration, the known information at the gateway is used to reduce the estimated frequency~\eqref{eqn:g:f} as
\begin{align}
{{\hat f}_{{n_{u,g}}}} = {\widetilde f_{{n_{u,g}}}} - {f_u} + {f_{{n_{{T_e}}}}} - {f_{{n_{d{l_e}}}}}\frac{{{\bf{v}}_{{n_{d{l_e}}}}^T{{\bf{k}}_{{n_{{{\left( {s,g} \right)}_e}}}}}}}{{{c_n}}} 
\label{eqn:f:sim}
\end{align}
where ${{\hat f}_{{n_{u,g}}}}$ is the reduced calibrated frequency of the $n$-th sample at the gateway and ${{\bf{k}}_{{n_{{{(u,s)}_e}}}}} = \frac{{{\bf{u}} - {{\bf{s}}_{{n_{u{l_e}}}}}}}{{\left\| {{\bf{u}} - {{\bf{s}}_{{n_{u{l_e}}}}}} \right\|}}$. The gateway uses~\eqref{eqn:f:sim} a long with the available data to build an analytical location-related equation as 
\begin{align}
{{\hat f}_{{n_{u,g}}}} = {f_u}\frac{{{\bf{v}}_{{n_{u{l_e}}}}^T{{\bf{k}}_{{n_{{{\left( {u,s} \right)}_e}}}}}}}{{{c_n}}} + {f_u}\frac{{{\bf{v}}_{{n_{u{l_e}}}}^T{{\bf{k}}_{{n_{{{\left( {u,s} \right)}_e}}}}}}}{{{c_n}}}\frac{{{\bf{v}}_{{n_{d{l_e}}}}^T{{\bf{k}}_{{n_{{{\left( {s,g} \right)}_e}}}}}}}{{{c_n}}}.
\label{eqn:foa:eq}
\end{align}
\begin{rmk}
	The value of $f_{T}$ changes for each sample due to the instability of satellite's electronics. Due to the difference between $f_u$ and $f_r$, the error of $f_T$ cannot be accurately derived, which reduces the localization accuracy. As a solution, we can use on-board spectrum monitoring~\cite{Christos:2015} to do on-board localization. Therefore, the sampled interference is not required to be down converted and thus its frequency is not influenced by the drift in the oscillator. Hence, the localization accuracy can be improved by on-board localization.   
\end{rmk}
In the following part, we describe the procedure to calculate the location of the interferer using the estimated and calibrated frequencies at the gateway.
\section{Localization Algorithm and Solution} \label{sec:alg}
The location-related equation in~\eqref{eqn:foa:eq} can be written as a function of the location of the unknown interferer as
\begin{align}
{f_n}\left( {\bf{u}} \right) = \frac{{{f_u}}}{{{c_n}}}\left[ {{\bf{v}}_{{n_{u{l_e}}}}^T{{\bf{k}}_{{n_{{{(u,s)}_e}}}}}} \right.\left. {\left( {1 + \frac{{{\bf{v}}_{{n_{d{l_e}}}}^T{{\bf{k}}_{{n_{{{(s,g)}_e}}}}}}}{{{c_n}}}} \right)} \right] - {\hat f_{{n_{u,g}}}}.
\label{eqn:f_u}
\end{align}
Since it is already known that the unknown interferer is located on the earth, at least two equations as in~\eqref{eqn:f_u} plus the equation of the earth surface are required to get an estimation for the location of the unknown interferer. To make a system of location-related equations, $N$ of the estimated frequencies at the gateway, with $N \ge 2$, are randomly selected. This system of nonlinear equations is solved using an iterative algorithm with the initial guess ${\bf{u}}_0$. To this end, the first-order Taylor series approximation around ${\bf{u}}_0$ is applied on each location-related equation to obtain
\begin{align}
{\bf{f}}\left( {\bf{u}} \right) \approx {\bf{f}}\left( {{{\bf{u}}_0}} \right) + {\bf{F}^{'}}\left( {{{\bf{u}}_0}} \right)\left( {{\bf{u}} - {{\bf{u}}_0}} \right),
\label{eqn:tay:app}
\end{align}
where ${\bf{f}}\left( {\bf{u}} \right) = \left[ {{f_1}\left( {\bf{u}} \right),...,{f_N}\left( {\bf{u}} \right),{{\left\| {\bf{u}} \right\|}^2} - {r^2}} \right]$, ${\left\| {\bf{u}} \right\|^2} = {r^2}$ is the surface of the earth equation, $r$ is the earth radius, ${\bf{f}}\left( {{{\bf{u}}_0}} \right) = \left[ {{f_1}\left( {{{\bf{u}}_0}} \right),...,{f_N}\left( {{{\bf{u}}_0}} \right),{{\left\| {{{\bf{u}}_0}} \right\|}^2} - {r^2}} \right]$, and ${\bf{F}^{'}}\left( {{{\bf{u}}_0}} \right)$ is the partial derivative matrix calculated at the initial guess as
\begin{align}
{\bf{F^{'}}}\left( {{{\bf{u}}_0}} \right) = \left[ {\begin{array}{*{20}{c}}
	{\frac{{\partial {f_{{1}}}\left( {{{\bf{u}}_0}} \right)}}{{\partial {u_1}}}}&{\frac{{\partial {f_{{1}}}\left( {{{\bf{u}}_0}} \right)}}{{\partial {u_2}}}}&{\frac{{\partial {f_{{1}}}\left( {{{\bf{u}}_0}} \right)}}{{\partial {u_3}}}}\\
	.&.&.\\
	.&.&.\\
	.&.&.\\
	{\frac{{\partial {f_{{N}}}\left( {{{\bf{u}}_0}} \right)}}{{\partial {u_1}}}}&{\frac{{\partial {f_{{N}}}\left( {{{\bf{u}}_0}} \right)}}{{\partial {u_2}}}}&{\frac{{\partial {f_{{N}}}\left( {{{\bf{u}}_0}} \right)}}{{\partial {u_3}}}}\\
	{2{u_1}}&{2{u_2}}&{2{u_3}}
	\end{array}} \right].
\label{eqn:der:def}
\end{align}
The partial derivatives of $f_n$ with respect to $u_m$ for $m=1,2,3$ are derived as $\frac{{\partial {f_n}\left( {\bf{u}} \right)}}{{\partial {u_m}}} = \frac{{{f_u}}}{{{c_n}}}{\eta _n}\left[ {{\bf{v}}_{{n_{u{l_e}}}}^T{{\bf{a}}_m}} \right]$ where
\begin{align}
&{{\bf{a}}_1} = \left[ {\begin{array}{*{20}{c}}
	{ - \frac{{{g_n} - {{\left( {{u_1} - {s_{{n_{{1_e}}}}}} \right)}^2}g_n^{ - 1}}}{{g_n^2}}}\\
	{\frac{{\left( {{u_1} - {s_{{n_{{1_e}}}}}} \right)\left( {{u_2} - {s_{{n_{{2_e}}}}}} \right)}}{{g_n^3}}}\\
	{\frac{{\left( {{u_1} - {s_{{n_{{1_e}}}}}} \right)\left( {{u_3} - {s_{{n_{{3_e}}}}}} \right)}}{{g_n^3}}}
	\end{array}} \right],
\label{eqn:a1}
\\
&{{\bf{a}}_2} = \left[ {\begin{array}{*{20}{c}}
	{\frac{{\left( {{u_2} - {s_{{n_{{2_e}}}}}} \right)\left( {{u_1} - {s_{{n_{{1_e}}}}}} \right)}}{{g_n^3}}}\\
	{ - \frac{{{g_n} - {{\left( {{u_2} - {s_{{n_{{2_e}}}}}} \right)}^2}g_n^{ - 1}}}{{g_n^2}}}\\
	{\frac{{\left( {{u_2} - {s_{{n_{{2_e}}}}}} \right)\left( {{u_3} - {s_{{n_{{3_e}}}}}} \right)}}{{g_n^3}}}
	\end{array}} \right],
\label{eqn:a2}
\\
&{{\bf{a}}_3} = \left[ {\begin{array}{*{20}{c}}
	{\frac{{\left( {{u_3} - {s_{{n_{{3_e}}}}}} \right)\left( {{u_1} - {s_{{n_{{1_e}}}}}} \right)}}{{g_n^3}}}\\
	{\frac{{\left( {{u_3} - {s_{{n_{{3_e}}}}}} \right)\left( {{u_2} - {s_{{n_{{2_e}}}}}} \right)}}{{g_n^3}}}\\
	{ - \frac{{{g_n} - {{\left( {{u_3} - {s_{{n_{{3_e}}}}}} \right)}^2}g_n^{ - 1}}}{{g_n^2}}}
	\end{array}} \right],
\label{eqn:a3}
\end{align}
${\eta _n} = \left( {1 + \frac{{{\bf{v}}_{{n_{d{l_e}}}}^T{{\bf{k}}_{{n_{{{(s,g)}_e}}}}}}}{{{c_n}}}} \right)$, ${g_n} = \left\| {{\bf{u}} - {{\bf{s}}_{{n_{u{l_e}}}}}} \right\|$, and ${{\bf{s}}_{{n_{u{l_e}}}}} = \left( {{s_{{n_{{1_e}}}}},{s_{{n_{{2_e}}}}},{s_{{n_{{3_e}}}}}} \right)$. We need to find the point ${\bf{u}} = {{\bf{u}}_1}$ to have ${\bf{f}}\left( {{{\bf{u}}_0}} \right) + {\bf{F^{'}}}\left( {{{\bf{u}}_0}} \right)\left( {{{\bf{u}}_1} - {{\bf{u}}_0}} \right)=0$ so that ${\bf{F^{'}}}\left( {{{\bf{u}}_0}} \right)\Delta {\bf{u}} =  - {\bf{f}}\left( {{{\bf{u}}_0}} \right)$, which is a system of linear equations with $\Delta {\bf{u}} = {{\bf{u}}_1} - {{\bf{u}}_0}$. After deriving $\Delta {\bf{u}}$, the initial guess is updated as
\begin{align}
{{\bf{u}}_{i+1}} = {{\bf{u}}_i} + \Delta {\bf{u}},
\label{eqn:itr}
\end{align}
and continues till $\left\| {\Delta {\bf{u}}} \right\| < \varepsilon$ where $\varepsilon$ depends on the required localization accuracy.
\section{Simulation Results}      \label{sec:sim}
\begin{table}[t]
	\caption{System Parameters} 
	\centering 
	\begin{tabular}{|l| l|}
		\hline
		Parameter & Value                            \\ [0.5ex]
		\hline
		\hline
	    Satellite orbit type                      & LEO, MEO, retro GEO, GEO   \\ 
		\hline
		Operating band                            & $K_a$ band \\
		\hline
		Uplink frequency, GHz                     &  30        \\
		\hline
		Satellite oscillator frequency, GHz       &  18        \\
		\hline
		Error bound for oscillator frequency, Hz  &  50        \\
		\hline
		Reference signal frequency, GHz           &  29        \\
		\hline
		Location of the gateway                   &  (5,14,0)  \\
		\hline
		Location of the unknown interferer        &  (30,20,0) \\
		\hline
	\end{tabular}
	\label{tab:sys:par}
\end{table}
\begin{figure}[]
	\centering
	\includegraphics[width=7cm]{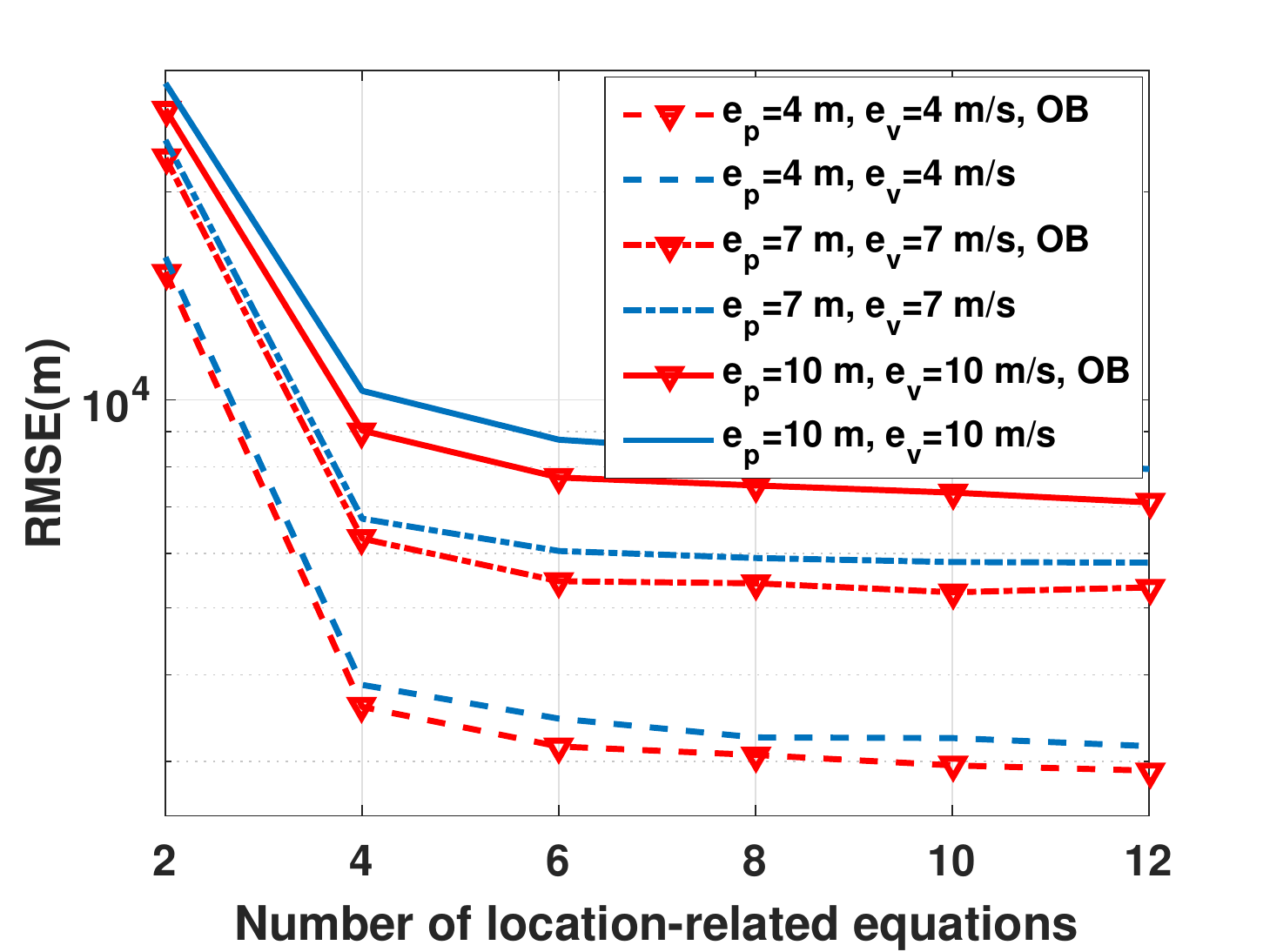}
	\caption{Localization RMSE versus the number of location-related equations when $\left\| {\bf{v}} \right\|=1544$ m/s, satellite altitude is $23000$ km, and the position of the reference transmitter is $(20,20,0)$.}
	\label{fig:rmse:eq:1}
\end{figure}
\begin{figure}[]
	\centering
	\includegraphics[width=7cm]{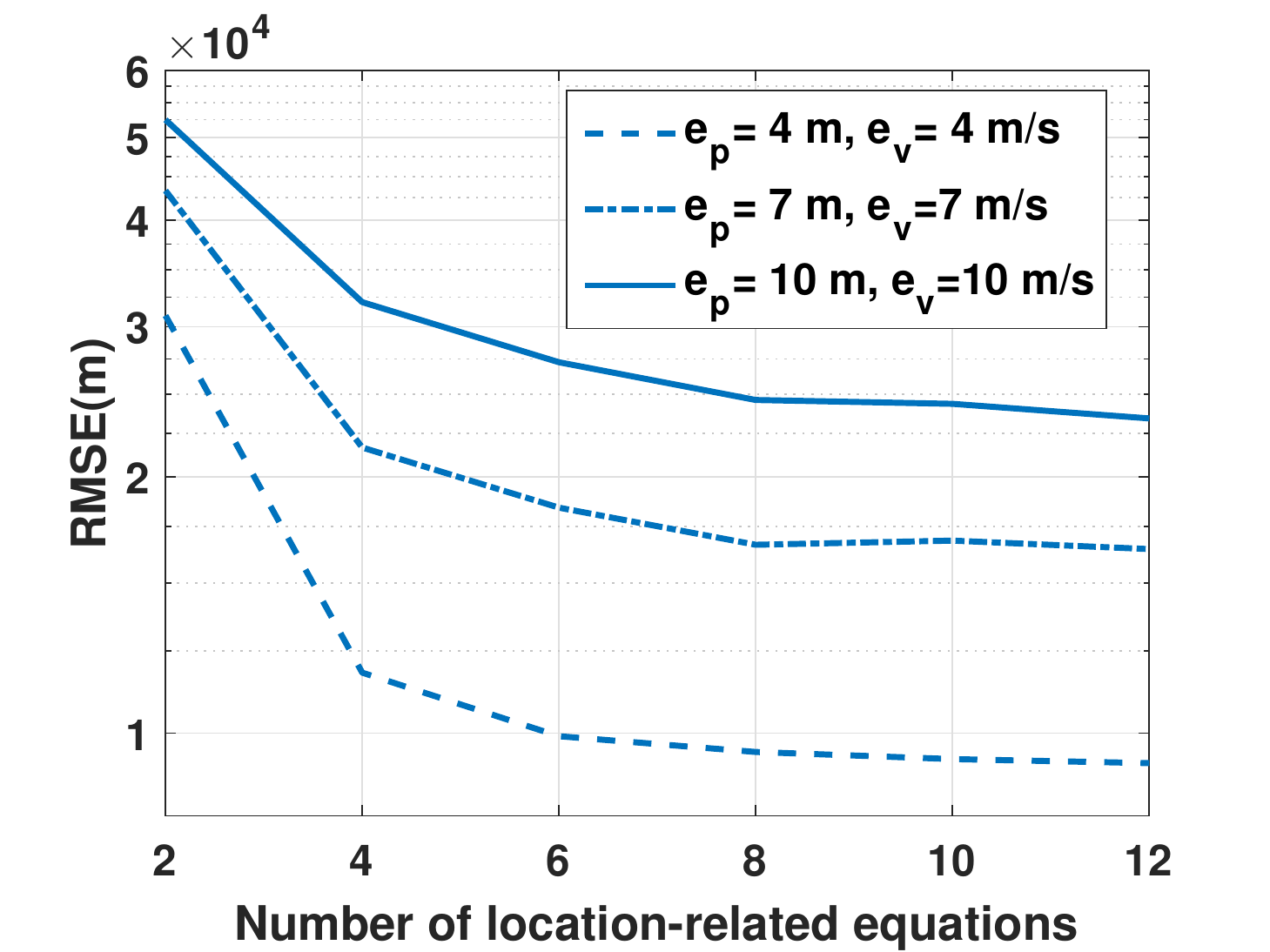}
	\caption{Localization RMSE versus the number of location-related equations when $\left\| {\bf{v}} \right\|=1544$ m/s, satellite altitude is $23000$ km, and the position of the reference transmitter is $(5,5,0)$.}
	\label{fig:rmse:eq:2}
\end{figure}
\begin{figure}[]
	\centering
	\includegraphics[width=8cm]{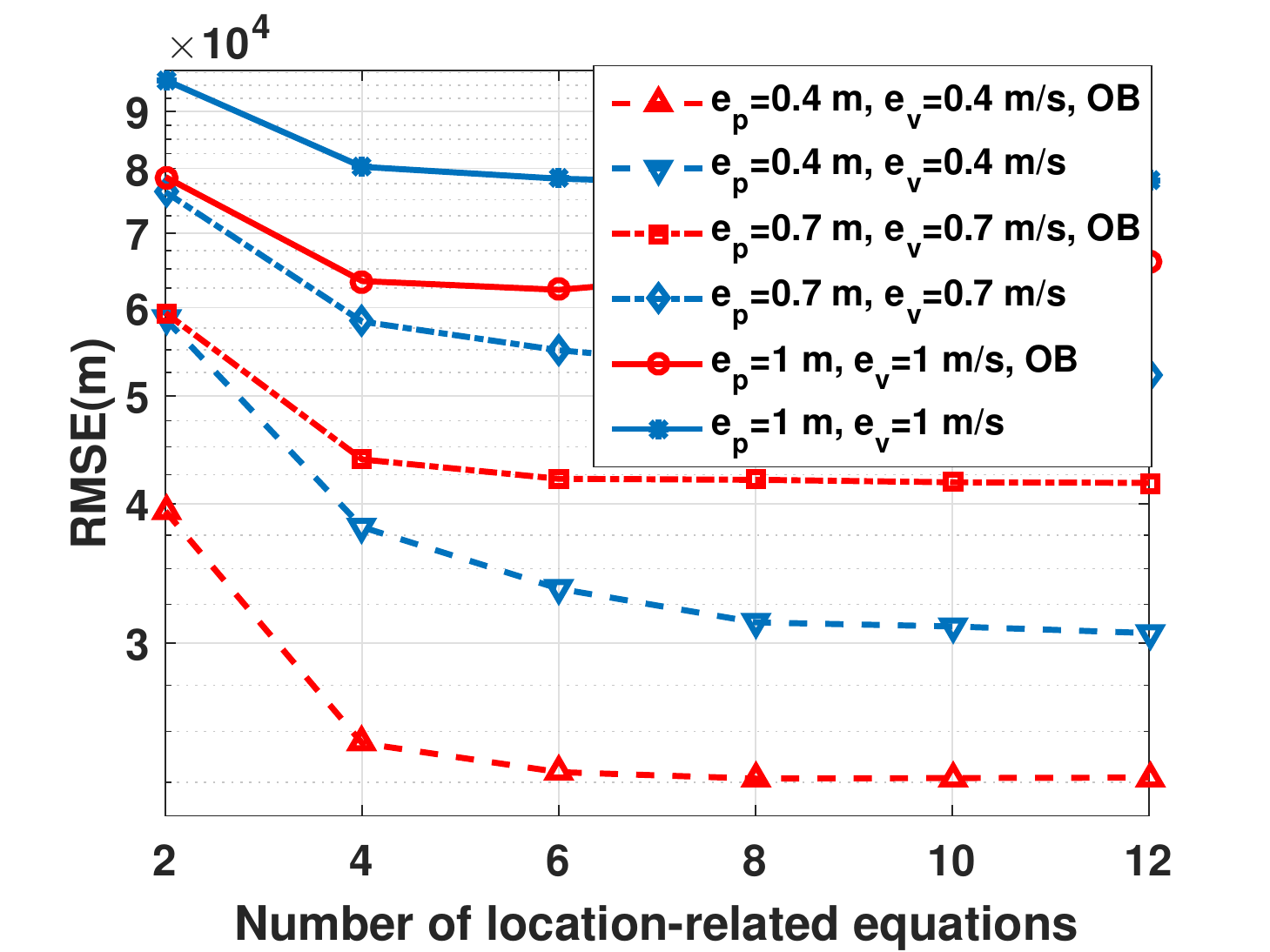}
	\caption{Localization RMSE versus the number of location-related equations for the GEO satellite when $\left\| {\bf{v}} \right\|=3.63$ m/s, satellite altitude is $35786$ km, and the position of the reference transmitter is $(20,20,0)$.}
	\label{fig:rmse:eq:geo}
\end{figure}
\begin{figure}[]
	\centering
	\includegraphics[width=7cm]{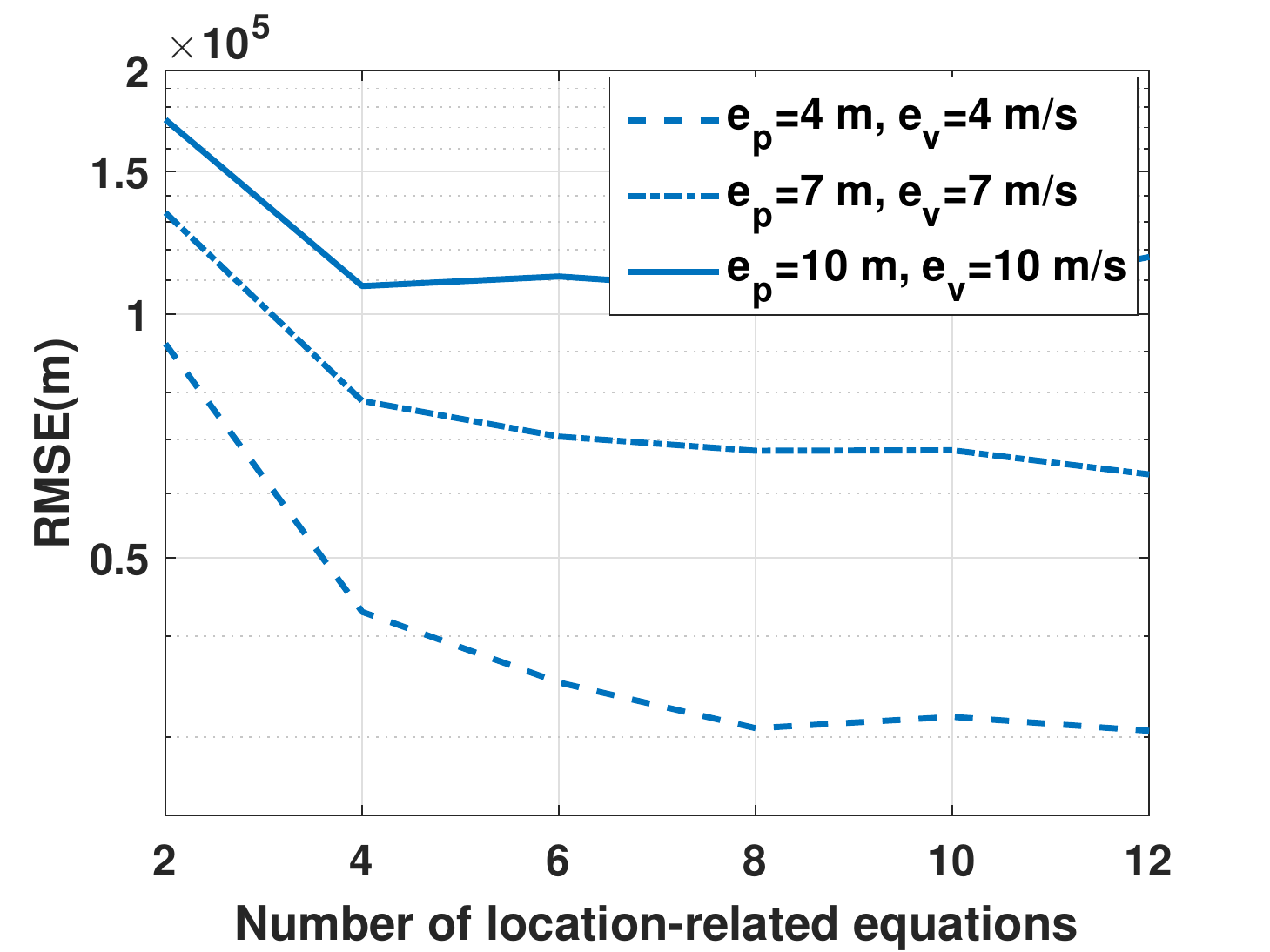}
	\caption{Localization RMSE versus the number of location-related equations using the FDOA technique of~\cite{Ho2013} when $\left\| {\bf{v}} \right\|=1544$ m/s, satellite altitude is $23000$ km, and the position of the reference transmitter is $(20,20,0)$.}
	\label{fig:rmse:fdoa}
\end{figure}
\begin{figure}[]
	\centering
	\includegraphics[width=7cm]{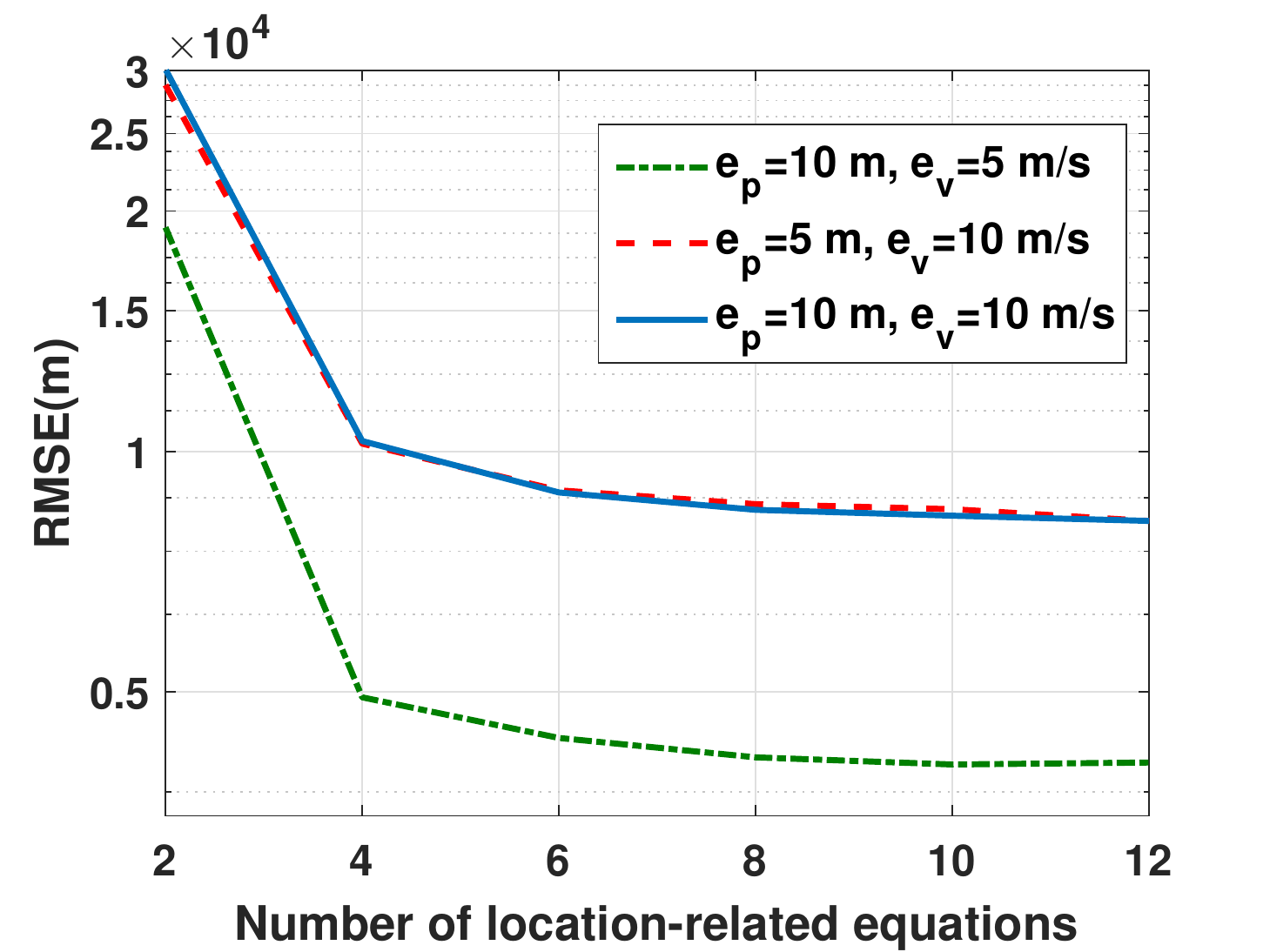}
	\caption{Comparing the effect of satellite's position and velocity errors on localization accuracy when $\left\| {\bf{v}} \right\|=1544$ m/s, satellite altitude is $23000$ km, and the position of the reference transmitter is $(20,20,0)$.}
	\label{fig:rmse:eq:dif:err}
\end{figure}
\begin{figure}[]
	\centering
	\includegraphics[width=7cm]{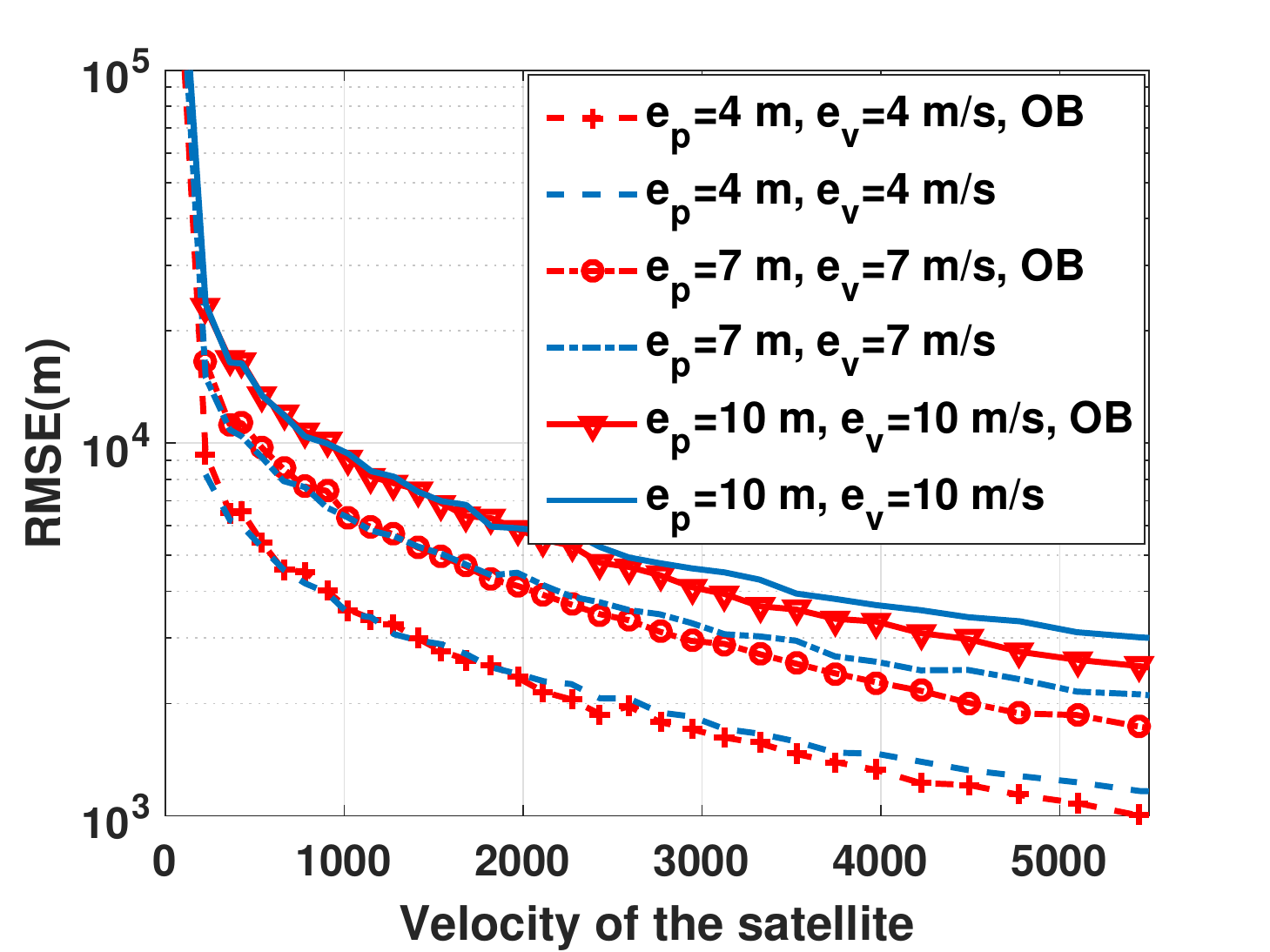}
	\caption{Localization RMSE versus the satellite velocity using six location-related equations when the position of the reference transmitter is $(20,20,0)$.}
	\label{fig:rmse:vel:1}
\end{figure}
\begin{figure}[]
	\centering
	\includegraphics[width=7cm]{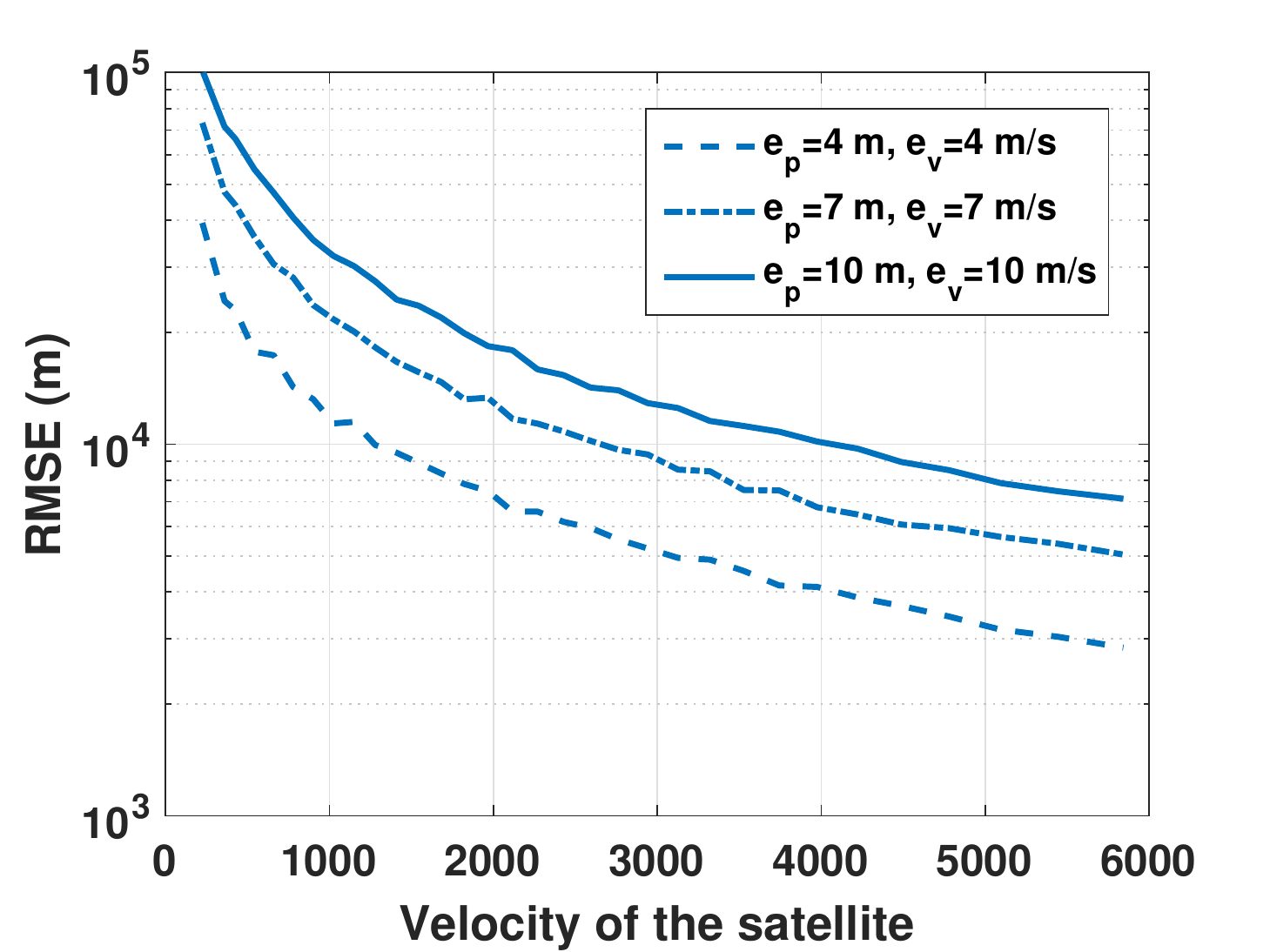}
	\caption{Localization RMSE versus the satellite velocity using six location-related equations when the position of the reference transmitter is $(5,5,0)$.}
	\label{fig:rmse:vel:2}
\end{figure}
In this section, we present different scenarios to evaluate the performance of the proposed localization technique. It is assumed that the processing at the satellite is quick enough so that the satellite positions and velocities can be considered to be the same during sampling and forwarding the interference and reference signals. Furthermore, the propagation speed of the electromagnetic wave is considered to be same for all frequency measurements. The locations of the system elements are shown by the Geographic coordinate system as (\textit{longitude},\textit{latitude},\textit{altitude}). The errors in the position and velocity of the satellite are shown by vectors ${{\bf{e}}_p}$ and ${{\bf{e}}_v}$, which their elements are uniform random variables within the distance $\left[ -{{e_{{p_{}}}},{e_{{p_{}}}}} \right]$ and $\left[ -{{e_{{v_{}}}},{e_{{v_{}}}}} \right]$, respectively. The acronym \textit{OB} is used instead of the term on-board in the legend of the figures to save space.  

For LEO, MEO, and retro GEO satellites, it is assumed that the satellite moves from ($0$,$0$,\textit{altitude}) to ($20$,$0$,\textit{altitude}) and samples the interference in every $0.5$ degree which results in 40 samples. Regarding the GEO satellite, it is assumed that the satellite collects 40 samples along a circular path with radius of $50$ km which takes one day to complete. The GEO satellite is located right above the intersection of zero degrees latitude and zero degrees longitude with the altitude $35786$ km. The rest of the parameters which are common for all the satellites are summarized in Table~\ref{tab:sys:par}.

For the first scenario, the effect of the number of location-related equations on the accuracy of the interferer localization is investigated in terms of the root mean square error (RMSE). The RMSE with respect to the number of location-related equations are shown in Figs.~\ref{fig:rmse:eq:1} and~\ref{fig:rmse:eq:2} for different positions of the reference signal. As it is seen, the localization accuracy improves by increasing the number of equations. Hence, we can figure out the required number of location-related equations to localize an unknown interferer within the required accuracy for specific amounts of system parameters and error bounds. By comparing Figs.~\ref{fig:rmse:eq:1} and~\ref{fig:rmse:eq:2}, we observe that if the location of the reference transmitter is closer to that of the unknown interferer, the localization accuracy increases. In addition, the localization RMSE with respect to the number of location-related equation for ob-board localization is presented in Fig.~\ref{fig:rmse:eq:1}. As seen, on-board localization further improves the localization accuracy since it avoids the error caused by the oscillator drift. The localization RMSE with respect to the number of location-related equations for the GEO satellite is presented in Fig.~\ref{fig:rmse:eq:geo}. Similar to Fig.~\ref{fig:rmse:eq:1}, it can be seen in Fig.~\ref{fig:rmse:eq:geo} that the localization accuracy improves by both increasing the number of location-related equations and using on-board localization. Since a GEO satellite moves relatively slow, the Doppler shift caused by its movement is small and can be easier influenced by the oscillator error. Hence, using on-board localization can considerably enhance the localization accuracy when a GEO satellite is sampling and forwarding the interference.

The localization RMSE with respect to the number of equations when using FDOA technique of~\cite{Ho2013} is presented in Fig.~\ref{fig:rmse:fdoa}. As compared with Fig.~\ref{fig:rmse:eq:1}, the RMSE considerably (by an order of magnitude) increases when FDOA is used instead of FoA. In FDOA approach, each equation is created by deducing two FoA measurements. Since FoAs can be close to each other, deducing them creates very small values which decreases the identifiability of the system of equations in~\eqref{eqn:tay:app}. 

In the next scenario, we analyze the sensitivity of the localization accuracy with respect to the errors in satellite's position and velocity. Localization RMSE with respect to location-related equations is presented in Fig.~\ref{fig:rmse:eq:dif:err} for different errors in position and velocity. As observed in Fig.~\ref{fig:rmse:eq:dif:err}, the localization accuracy is much more sensitive to velocity errors than the position errors. This is due to the fact that each error in satellite's position is divided by ${\left\| {{\bf{u}} - {{\bf{s}}_{{n_{u{l_e}}}}}} \right\|}$ and ${\left\| {{{\bf{s}}_{gw}} - {{\bf{s}}_{{n_{d{l_e}}}}}} \right\|}$ in~\eqref{eqn:f_u}, which reduces the effect of the position error.

In the last scenario, the effect of the satellite velocity on the localization accuracy is investigated. The localization RMSE with respect to the satellite's velocity is shown in Figs.~\ref{fig:rmse:vel:1} and~\ref{fig:rmse:vel:2}. The results show that the localization accuracy is improved when the velocity of the satellite increases. Similar to the first scenario, if the locations of the reference and unknown transmitters are closer to each other, the localization accuracy increases considerably. Furthermore, Fig.~\ref{fig:rmse:vel:1} shows that the localization accuracy is improved using on-board approach.   
\section{Conclusion}  \label{sec:con}
We proposed using the FoA technique to localize an unknown interferer while only relying on either the affected satellite, or the satellite dedicated to interference localization. We used a reference signal to calibrate the estimated frequency of the interferer at the gateway, and built location-related equations using the values of satellite's oscillator frequency, velocities, and positions. It was shown that increasing the number of location-related equations, i.e. measurements, can improve the localization accuracy. In addition, the localization accuracy improved when the affected satellite had a higher velocity. The results showed that a closer reference transmitter to the location of the unknown interferer enhances the localization accuracy. Moreover, the simulations showed that using the proposed on-board localization approach can further enhance the localization accuracy since the oscillator error is avoided, particularly for on-board GEO localization. It was observed that using FoA technique instead of FDOA improves the localization accuracy considerably since it increases the identifiability of the system of location-related equations.    
\section*{Acknowledgment} 
This work was supported by the National Research Fund (FNR) of Luxembourg under AFR grant  for the project ``Physical Layer Security in Satellite Communications (ref. 5798109)'', SeMIGod, and SATSENT.
\bibliographystyle{IEEEtran}


\begin{thebibliography}{10}
	\providecommand{\url}[1]{#1}
	\csname url@samestyle\endcsname
	\providecommand{\newblock}{\relax}
	\providecommand{\bibinfo}[2]{#2}
	\providecommand{\BIBentrySTDinterwordspacing}{\spaceskip=0pt\relax}
	\providecommand{\BIBentryALTinterwordstretchfactor}{4}
	\providecommand{\BIBentryALTinterwordspacing}{\spaceskip=\fontdimen2\font plus
		\BIBentryALTinterwordstretchfactor\fontdimen3\font minus
		\fontdimen4\font\relax}
	\providecommand{\BIBforeignlanguage}[2]{{%
			\expandafter\ifx\csname l@#1\endcsname\relax
			\typeout{** WARNING: IEEEtran.bst: No hyphenation pattern has been}%
			\typeout{** loaded for the language `#1'. Using the pattern for}%
			\typeout{** the default language instead.}%
			\else
			\language=\csname l@#1\endcsname
			\fi
			#2}}
	\providecommand{\BIBdecl}{\relax}
	\BIBdecl
	
	\bibitem{Smith:1991}
	J.~Smith, W.W. and P.~Steffes, ``A satellite interference location system using
	differential time and phase measurement techniques,'' \emph{IEEE Aerosp.
		Electron. Syst. Mag.}, vol.~6, no.~3, pp. 3--7, Mar. 1991.
	
	\bibitem{Ho:1993}
	K.~Ho and Y.~Chan, ``Solution and performance analysis of geolocation by
	{TDOA},'' \emph{IEEE Trans. Aerosp. Electron. Syst.}, vol.~29, no.~4, pp.
	1311--1322, Oct. 1993.
	
	\bibitem{Haworth:1997}
	D.~P. Haworth, N.~G. Smith, R.~Bardelli, and T.~Clement, ``Interference
	localization for {EUTELSAT} satellites--the first european transmitter
	location system,'' \emph{Int. J. Satell. Commun.}, vol.~15, no.~4, pp.
	155--183, Jul. 1997.
	
	\bibitem{Ho:1997}
	K.~Ho and Y.~Chan, ``Geolocation of a known altitude object from tdoa and fdoa
	measurements,'' \emph{IEEE Trans. Aerosp. Electron. Syst.}, vol.~33, no.~3,
	pp. 770--783, Jul. 1997.
	
	\bibitem{Pattison:2000}
	T.~Pattison and S.~Chou, ``Sensitivity analysis of dual-satellite
	geolocation,'' \emph{IEEE Trans. Aerosp. Electron. Syst.}, vol.~36, no.~1,
	pp. 56--71, Jan. 2000.
	
	\bibitem{Teng:2010}
	T.~Li, F.~Guo, and W.~Jiang, ``A novel emitter localization method using an
	interferometer on a spin-stabilized satellite,'' in \emph{Wireless
		Communications and Signal Processing (WCSP)}, Suzhou, China, Oct. 2010.
	
	\bibitem{Jinzhou:2012}
	J.~Li, F.~Guo, and W.~Jiang, ``A linear-correction least-squares approach for
	geolocation using {FDOA} measurements only,'' \emph{Chinese Journal of
		Aeronautics}, vol.~25, no.~5, pp. 709--714, May 2012.
	
	\bibitem{Effland:1991}
	\BIBentryALTinterwordspacing
	J.~Effland, J.~Gipson, D.~Shaffer, and J.~Webber, ``Method and system for
	locating an unknown transmitter,'' Apr. 1991, {US} Patent 5,008,679.
	[Online]. Available: \url{http://www.google.com/patents/US5008679}
	\BIBentrySTDinterwordspacing
	
	\bibitem{Haworth:2000}
	\BIBentryALTinterwordspacing
	D.~Haworth, ``Locating the source of an unknown signal,'' Jan. 2000, {US}
	Patent 6,018,312. [Online]. Available:
	\url{http://www.google.com/patents/US6018312}
	\BIBentrySTDinterwordspacing
	
	\bibitem{Rideout:2004}
	\BIBentryALTinterwordspacing
	R.~Rideout, P.~Edmonds, S.~Duck, D.~Haworth, and C.~Griffin, ``Method and
	apparatus for locating the source of unknown signal,'' Jan. 2004, {US} Patent
	6,677,893. [Online]. Available: \url{http://www.google.com/patents/US6677893}
	\BIBentrySTDinterwordspacing
	
	\bibitem{Ho:2010:1}
	\BIBentryALTinterwordspacing
	D.~Ho, J.~Chu, and M.~Downey, ``Determining a geolocation solution of an
	emitter on earth based on weighted least-squares estimation,'' Feb. 2010,
	{US} Patent 7,663,547. [Online]. Available:
	\url{http://www.google.com.br/patents/US7663547}
	\BIBentrySTDinterwordspacing
	
	\bibitem{Ho:2010:2}
	\BIBentryALTinterwordspacing
	------, ``Determining a geolocation solution of an emitter on earth using
	satellite signals,'' Feb. 2010, {US} Patent 7,667,640. [Online]. Available:
	\url{http://www.google.com.na/patents/US7667640}
	\BIBentrySTDinterwordspacing
	
	\bibitem{Ho2013}
	\BIBentryALTinterwordspacing
	------, ``Determining transmit location of an emitter using a single
	geostationary satellite,'' Jun. 2013, {US} Patent 8,462,044. [Online].
	Available: \url{http://www.google.com/patents/US8462044}
	\BIBentrySTDinterwordspacing
	
	\bibitem{Ha:1987}
	T.~Ha and R.~Robertson, ``Geostationary satellite navigation systems,''
	\emph{IEEE Trans. Aerosp. Electron. Syst.}, vol.~23, no.~2, pp. 247--254,
	Mar. 1987.
	
	\bibitem{Christos:2015}
	C.~Politis, S.~Maleki, S.~Chatzinotas, and B.~Ottersten, ``Harmful interference
	threshold and energy dector for on-board interference detection,'' submitted
	to Vechicular Technology conference (VTC), 2016.
	
\end{thebibliography}
\end{document}